# Correlating activity and defects in (photo)electrocatalysts using in-*situ* transient optical microscopy

[1,2,3]Camilo A. Mesa, [1,4,5]Michael Sachs, [2,6]Ernest Pastor, [7]Nicolas Gauriot, [7]Alice J. Merryweather, [8]Miguel A. Gomez-Gonzalez, [8]Konstantin Ignatyev, [2]Sixto Giménez, [7]Akshay Rao, [1,9]James R. Durrant and [7,10]Raj Pandya✉

[1]Department of Chemistry and Centre for Processable Electronics, Imperial College London, London, W12 0BZ, United Kingdom

[2]Institute of Advanced Materials (INAM) Universitat Jaume I 12006 Castelló Spain

[3]Sociedad de Doctores e Investigadores de Colombia, Grupo de Investigación y Desarrollo en Ciencia Tecnología e Innovación - BioGRID, Bogotá 111011, Colombia

[4]SLAC National Accelerator Laboratory, Menlo Park, CA, USA

[5]PULSE Institute, SLAC National Accelerator Laboratory, Stanford University, Stanford, CA, USA

[6]IPR–Institut de Physique de Rennes, CNRS-Centre national de la recherche scientifique, UMR Université de Rennes, 35000 Rennes, France

[7]Cavendish Laboratory, University of Cambridge, J.J. Thomson Avenue, CB3 0HE, Cambridge, UK

[8]Diamond Light Source Ltd., Harwell Science and Innovation Campus, Didcot, Oxfordshire, OX11 0DE, United Kingdom

[9]SPECIFIC IKC, College of Engineering, Swansea University, Swansea, SA2 7AX, United Kingdom

[10]Laboratoire Kastler Brossel, ENS-Université PSL, CNRS, Sorbonne Université, Collège de France, 24 rue Lhomond, 75005 Paris, France

✉Correspondance to: rp558@cam.ac.uk

**Abstract**

(Photo)electrocatalysts capture sunlight and use it to drive chemical reactions such as water splitting to produce $H_2$. A major factor limiting photocatalyst development is their large heterogeneity which spatially modulates reactivity and precludes establishing robust structure-function relationships. To make such links requires simultaneously probing of the electrochemical environment at microscopic length scales (nm to μm) and broad timescales (ns to s). Here, we address this challenge by developing and applying in-*situ* steady-state and transient optical microscopies to directly map and correlate local electrochemical activity with hole lifetimes, oxygen vacancy concentration and the photoelectrode's crystal structure. Using this combined approach alongside spatially resolved X-Ray absorption measurements, we study microstructural and point defects in prototypical hematite (α-$Fe_2O_3$) photoanodes. We demonstrate that regions of α-$Fe_2O_3$, adjacent to microstructural cracks have a better photoelectrochemical response and reduced back electron recombination due to an optimal oxide vacancy concentration, with the film thickness and carbon impurities also dramatically influencing activity in a complex manner. Our work highlights the importance of microscopic mapping to understand activity and the impact of defects in even, seemingly, homogeneous solid-state metal oxide photoelectrodes.

**Main**

The (photo)electrocatalytic splitting of water to hydrogen and oxygen is a key reaction for renewable energy conversion and storage. Metal-oxide semiconductor photoanodes such as $TiO_2$, $WO_3$ and $\alpha$-$Fe_2O_3$, continue to attract great interest for water splitting despite exhibiting large overpotentials[1–4], due to their high chemical stability, low cost, optoelectronic properties and relative ease for scale[5]. The efficiency of catalytic reactions in these materials is primarily determined by their ability to generate and efficiently separate charge carriers, and the kinetics of water oxidation when these charges accumulate at the material surface. It is the kinetic mismatch between these charge generation and accumulation processes, in particular the bulk carrier lifetimes (*i.e.* recombination) and catalysis (*i.e.* hole transfer to water), over a range of microscopic length scales and reaction timescales[6,7], that limits the overall reaction efficiency in these metal-oxide photoanodes[8–10].

Structural and chemical disorder generated during the material synthesis[11–13] can strongly impact the photogenerated hole dynamics in metal-oxide semiconductors. This disorder can be broadly split into two categories: (i) micro-structural defects such as cracks and grain boundaries[14–17] and (ii) 0D point defects, which might include impurity doping or composition heterogeneity such as oxygen vacancies[18,19]. Both are known to play important roles in the catalytic properties of photoelectrodes, but investigating their function in-*situ* with the necessary simultaneous chemical, spatial and temporal resolution has remained challenging.

Techniques such as scanning electrochemical microscopy (SECM) have been used to visualise (with nanometre resolution) the distribution of dopants on photoelectrode surfaces and study water splitting in heterojunction photoelectrodes[20–24]. However, SECM, and its' variants, are challenging to apply on rough electrode surfaces and generally provide limited temporal information on carrier dynamics[25,26]. Alternatively, a combination of scanning electron microscopy[27,28], surface photovoltage microscopy[15,16,29,30] and Kelvin probe measurements have been used to link defects and activity[31] and visualise charge transfer at a range of photocatalytic surfaces and interfaces, but *in situ* studies remain challenging[32,33]. Similarly, (super-resolution) fluorescence microscopy[34] has enabled probing the influence of inter-facet junctions[35,36] and visualize proton-transfer within zeolite crystals[17]. But the reliance on indirect exogenous fluorescent probes makes this method difficult to apply to a wide-range of catalytic systems.

A need hence remains to develop and apply, inexpensive, non-destructive and label-free tools for investigation of (photo)electrocatalysts as they perform water splitting. Furthermore, given the wide-range of microstructural defects influencing water oxidation kinetics/catalysis, an approach with chemical, temporal and spatial resolution must be taken such that a complete picture can be obtained.

In this work, we develop a multi-modal microscopy setup to map, in a correlated manner, with sub-micron resolution, the surface of photoelectrodes using (spectro)electrochemical, transient reflection (ns to ms time range), and Raman spectroscopies. We apply this tool to image the local photocurrent across different regions of a an $\alpha$-$Fe_2O_3$ photoanode and correlate it with both the population dynamics of photogenerated holes, the local oxide vacancy concentration, and structure. We find that, even in relatively homogeneous thin-film photoelectrodes, microscopic defects strongly influence hole dynamics and local activity, primarily *via* the rate of unfavourable back electron recombination reactions which impact the 'local' electrode J-V fill-factor. We further develop our findings using *in-situ* X-Ray near-edge absorption microspectroscopy (µ-XANES) providing mapping of oxygen vacancies, and their effect on the accumulation of high-valent, catalytically active species, in areas of the photoanodes. Our results emphasize the importance of moving beyond an ensemble approach to build structure-function relationships in these photocatalytic systems.

## Correlated optical microscopy

In this work, we study ~300 nm thick films of α-Fe$_2$O$_3$ (hematite) prepared using a hydrothermal synthesis method (see **Methods**) reported elsewhere[37,38]. A modified electrochemical cell (**Figure 1**) with optical access is used to perform microspectroscopy under bias and measure local photoelectrochemical activity. All the applied potentials are referenced to the reversible hydrogen electrode (RHE; $V_{RHE}$) and, unless otherwise indicated 0.1 M NaOH (aq; pH 13.6) was used as an electrolyte.

Photocurrent mapping as shown in the top panel of **Figure 2** builds a spatial picture of the catalytic activity, *via* the photocurrent generated when illuminating with 532 nm light specific regions of the sample. The photocurrent density is correlated with the hole lifetime using transient (diffuse) reflection microscopy (TRM). Here, a combination of a focussed pump pulse at 400 nm and focussed CW probe pulse at 638 nm allows monitoring of the time resolved photoinduced absorption of holes[39] (see **Methods** for further details). In this way, we measure the population dynamics of photogenerated holes with sub-micron spatial precision over 10 μm × 10 μm regions as a function of applied bias (**Figure 2**, middle panel). Over the same regions spectrally-resolved, bias dependent, microscopic reflection imaging measurements are performed as shown in the bottom panel of **Figure 2**. The magnitude and bias sensitivity of the reflectivity are correlated with the oxygen vacancy population[40,41] *via* intraband absorptions, as discussed below. Finally, Raman imaging (bottom panel of **Figure 2**) provides insight into the local chemical structure of the regions mapped and their evolution under bias.

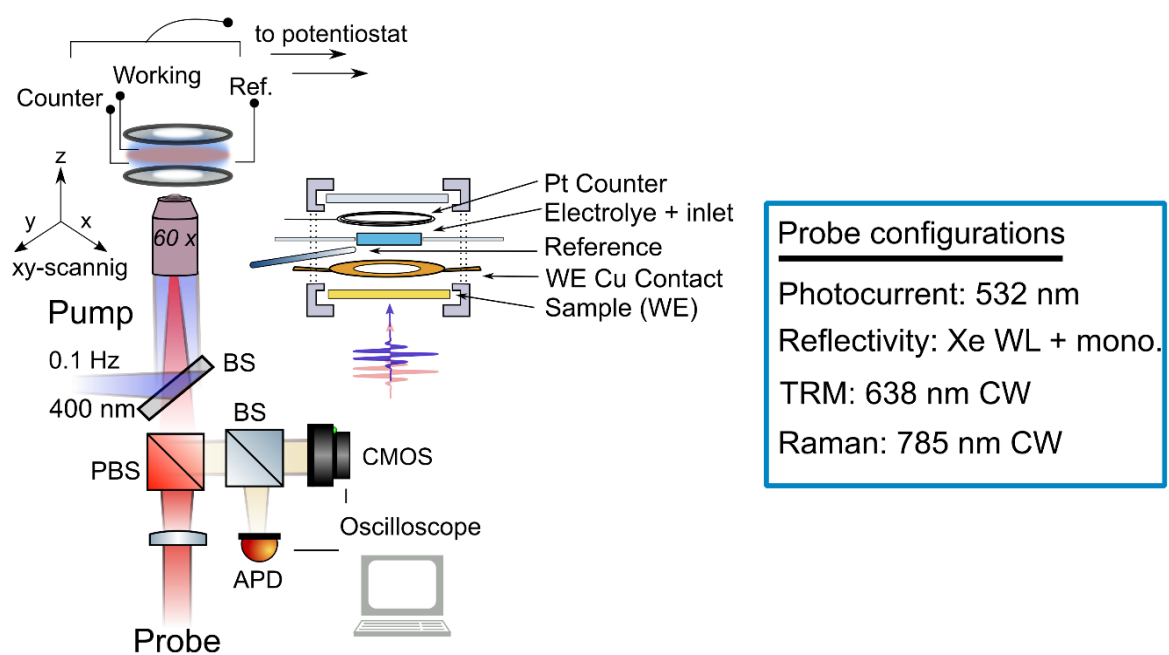

**Figure 1: Schematic of multi-modal correlated microscopy setup:** Pump and probe sources are coupled into a home-built reflection microscope. The pump (only used in transient reflection microscopy (TRM)) is a 0.1 Hz, 400 nm pulsed laser (5 ns pulse width), focussed to 440 ± 10 nm on the sample with a 60 ×, 0.9 N.A. objective. The probe is varied depending on the experiment as detailed in the blue box (size range: 900 ± 10 nm to 1.3 ± 0.05 μm spot size). In the case of reflectivity measurements, the probe is wide-field (40 μm spot size). An 80:20 beam splitter, diverts signals to an avalanche photodiode (APD) for collection and camera for imaging. A 3-electrode electrochemical cell with optical access allows for application of potentials and collection of spatio-temporal optical signals. In experiments with focussed light, the sample is scanned to build 2D images.

## Local Photocatalytic Activity

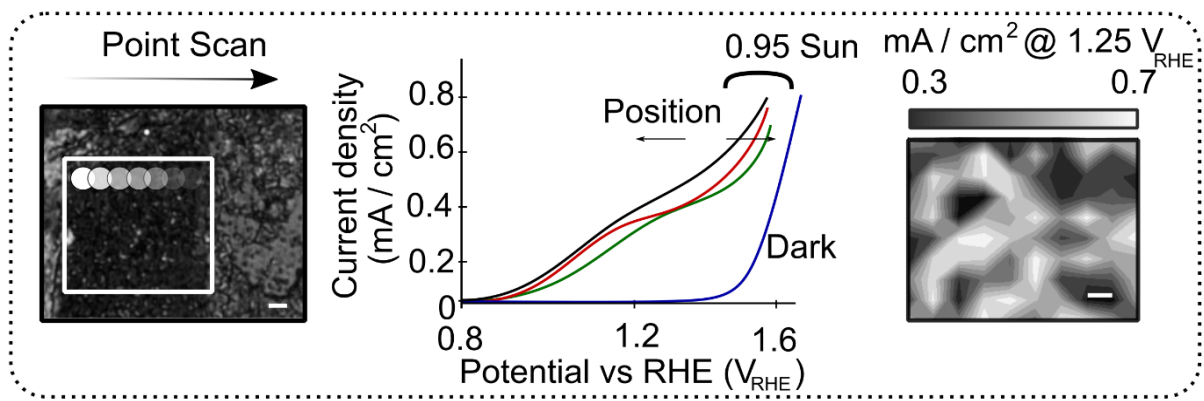

## Microscopic Hole Dynamics

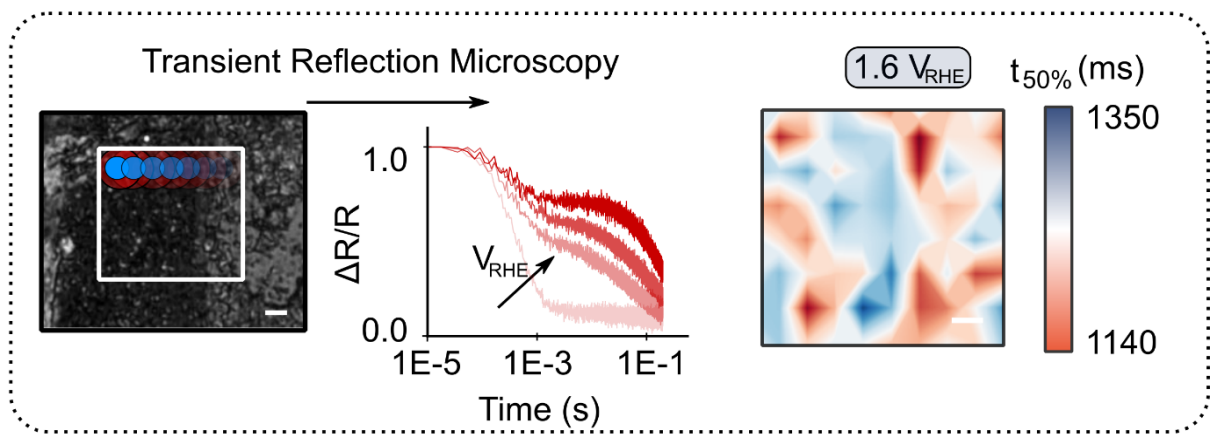

## Vibrational Structure & Oxide Vaccancies

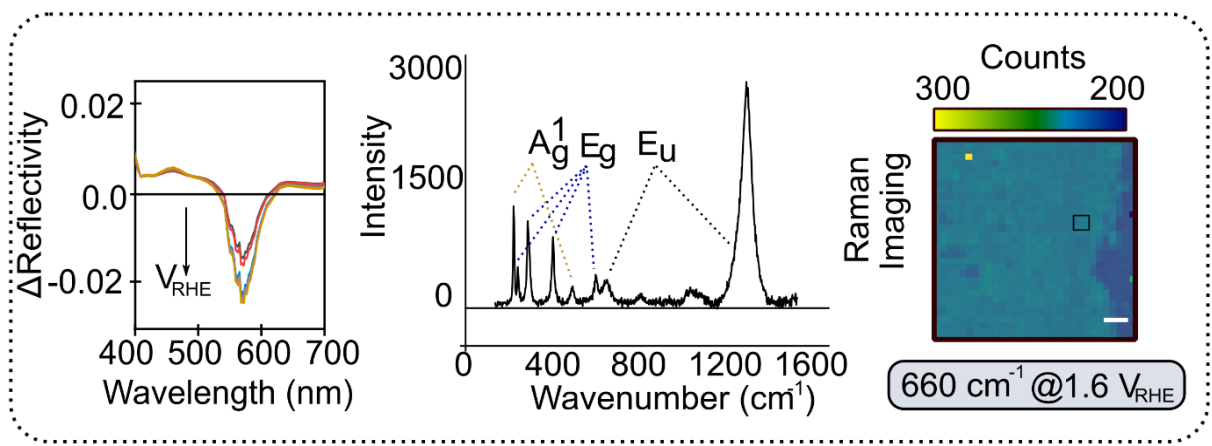

**Figure 2:** *In-situ* **simultaneous mapping of activity, hole dynamics and defect composition in a model (photo)electrocatalyst.** **(Top)** Scanning a focussed 532 nm light source (~1.3 μm FWHM; ~8 MW cm$^{-2}$ intensity at sample) allows the photocurrent density under illumination to be measured locally at different regions in the sample. The spatial photocurrent density is then plotted at 1.25 V vs RHE ($V_{RHE}$). A dark CV is shown in the middle panel (in blue) for reference. **(Middle)** Scanning of the sample allows transient reflection microscopy kinetics (ΔR/R) to be obtained at each spatial location. The kinetics from each TRM experiment can be used to obtain the time taken for the signal to decay to half its initial value ($t_{50\%}$) as a function of bias and spatial location. **(Bottom)** Differential reflection spectrum as a function of bias (with respect to open circuit potential), shown for the region marked by the black box in the middle panel. Raman imaging (500 ± 10 nm spatial resolution) of a hematite photoelectrode measured using point scanning. The Raman spectrum of α-Fe$_2$O$_3$ is plotted in the bottom left with the various modes and their corresponding symmetries (A$^1_g$, E$_g$ and E$_u$) marked. Image displays the spatial intensity of the 660 cm$^{-1}$ mode at open circuit potential. These measurements give spatial insight into the concentration of oxygen vacancies. Scale bar on all images in the figure are 1 μm.

## (Photo)electrocatalysis at microstructural breaks in the film

Microstructural defects in (photo)electrocatalysts often occur in the form of breaks and cracks. In the hematite films studied here these range on the order of 500 nm to 1500 nm in width (see **Supplementary Note 1**). We sub-categorise a prototypical 10 × 10 um region containing cracks, as shown in the optical and SEM images in **Figure 3a**, into three regions of similar thickness: crack (termed CR; marked with blue asterisk), adjacent-to-crack (termed ACR; marked with red/green asterisk) and flat un-cracked film (termed FL; marked with orange asterisk). Plotting the EDX spectra of the ACR and CR regions subtracted from that in the FL regions (**Figure 3a** right panel), shows the CR regions have a lower Fe and O content (peaks at 0.71 eV/6.4 eV and 0.51 eV, respectively) as compared to FL (remaining peaks arise from trace impurities Si and Cl, likely arising from the synthesis, and Sn from the FTO substrate underneath; see **Supplementary Note 2**). Conversely, ACR regions are richer in Fe than FL regions but contain a similar amount of O as the latter. Indeed, the Fe:O ratio in ACR regions is lower than in FL regions (0.59 *versus* 0.65, respectively), indicating that these regions are oxygen deficient (see further discussion below)[42,43].

To link the composition with catalytic activity we measure the photocurrent over the same area as shown in **Figure 3b** (see **Methods**). Two salient features can be discerned from the photocurrent density map (plotted at 1.3 V$_{RHE}$) and curves: (i) at the maximum potential ~1.5 V$_{RHE}$ all three regions produce a similar amount of photocurrent and (ii) the oxygen deficient ACR region exhibit a photocurrent onset ~200 mV earlier than the FL region (~0.7 V$_{RHE}$ compared to ~0.9 V$_{RHE}$, respectively). Below the saturation potential both water oxidation and back electron transfer reactions (BER) occur. From the curves in the **Figure 3b** the earlier photocurrent onset for ACR suggests BER is turned off more quickly here and ACR regions have a higher performance (fill factor). We note these trends in the photocurrent density are also seen in other regions of the film near cracks (see **Supplementary Note 3**).

To further understand these performance differences, we study the dynamics of photogenerated holes, over the same spatial region as imaged in **Figure 3a**, using transient reflection microscopy. Data is collected probing at 638 nm, previously assigned to the absorption of long lived hematite holes[44]. Although several equations can be used to describe the transient decays (ΔR/R) we focus on using $t_{50\%}$ (half-life) to assay the effective decay time and avoid multiple fitting parameters. The rate of decay represents a combination of the primarily bulk electron-hole recombination processes and slower decay processes of surface localised holes. This second long-lived component at mild oxidative potentials, will have contributions from both water oxidation and processes such as back electron transfer that affect the surface hole population[45,46]. In **Figure 3c** we plot ΔR/R at ACR, CR and FL regions as highlighted in **Figure 3a**. For all three regions $t_{50\%}$ increases with bias, *i.e.* slower decays consistent with previous observations, but as the potential is raised $t_{50\%}$ follows the trend ACR ≈ FL > CR. At strong biases (~1.6 V$_{RHE}$) $t_{50\%}$ is ~1110 ms in ACR and FL, but ~940 ms in CR, whereas at lower biases (0.8 V$_{RHE}$), where BER quickens the hole decay, $t_{50\%}$, is ~115 ms and ~86 ms for the ACR/FL and CR regions, respectively. The results in **Figure 3c** suggest that more potential is required to separate charges and 'turn off' BER in the crack regions, in keeping with the higher potential needed for generation of photocurrent. We note that if we examine the magnitude of the non-normalised kinetics as shown in **Figure 3d** (and **Supplementary Note 4**) we find that the ΔR/R signal strength follows the order ACR > FL > CR. This quantity will be a measure of yield of surface holes suggesting there are fewer holes generated in the cracks.

To test the above hypotheses further we examine the oxygen vacancy density which has been suggested to correlate with back electron transfer reactions. The spatially dependent reflectivity is measured in **Figure 4a** as a function of a bias (with respect to the reflectivity at 0.5 $V_{RHE}$) from the FL, CR and ACR regions. In FL and CR there is a single negative peak in the (non-normalised) differential reflection spectrum centred at ~570 nm, in ACR there are two peaks, one at 550 nm and another at 575 nm. The differential reflectivity signal at 570 nm has been previously assigned to the generation of ionised oxygen vacancies (OVs), *i.e.* electron removal, corresponding to space charge layer formation, although reactive surface states may also play a role in this signal. Such OVs are typically associated with an excess of electrons resulting in the reduction of $Fe^{3+}$ to $Fe^{2+}$ in α-$Fe_2O_3$. The larger magnitude of the differential reflectivity in ACR regions as compared to FL/CR suggests that the vacancy concentration is higher. The presence of the second peak at 550 nm in the reflectivity spectrum in ACR regions is challenging to assign. Based on the fact this second peak is more energetic, and therefore located more positive into the valence band of the semiconductor it may be related to a second type of hole species as hypothesised elsewhere[47,48]. It is remarked that in other metal oxides, such as $WO_3$, an intermediate oxide vacancy concentration of 2 to 3% has been found to be optimal for water splitting[41]. This was suggested to result from the vacancies aiding local charge separation and releasing charge carriers to the surrounding lattice for conductivity. The higher [OV] in ACR regions, as compared to CR (and even FL), alongside the larger fill factor and greater $t_{50\%}$ of ACR suggests that the oxygen vacancy concentration here sits in an intermediate regime where the negative trapping effects are balanced against the charge separation properties they imbue.

To gain further understanding of the general effect of applied potential on oxygen vacancies we perform XANES measurements on homogeneous regions of hematite film (sample synthesised as above). Sun and co-workers[49] have reported that the relative intensity of the Fourier transformed XANES data in R-space for the Fe–O bond can be indicative of the relative density of such oxygen vacancies in hematite films. **Figure 4b** shows the R-space data, from the XANES spectra (see **Supplementary Note 5**), where the relative intensity of the Fe–O bond (at 1.5 Å) appears to be insensitive to the strong positive bias. On first glance this suggests that the density of such OVs does not change under a positive bias in dark conditions. However, the change in the intensity of the Fe–Fe bond (at 2.48 Å) on applying a potential of 1.5 V vs RHE, indicates that the introduction of disorder by application of the positive potential. Such disorder can tentatively be assigned to OVs oxidation correlating broadly with the behaviour observed in the differential reflectivity data shown in **Figure 4a** where applying a positive potential reduces the concentration of filled OVs (resulting in $Fe^{2+}$ to $Fe^{3+}$ oxidation and a decrease in reflectivity).

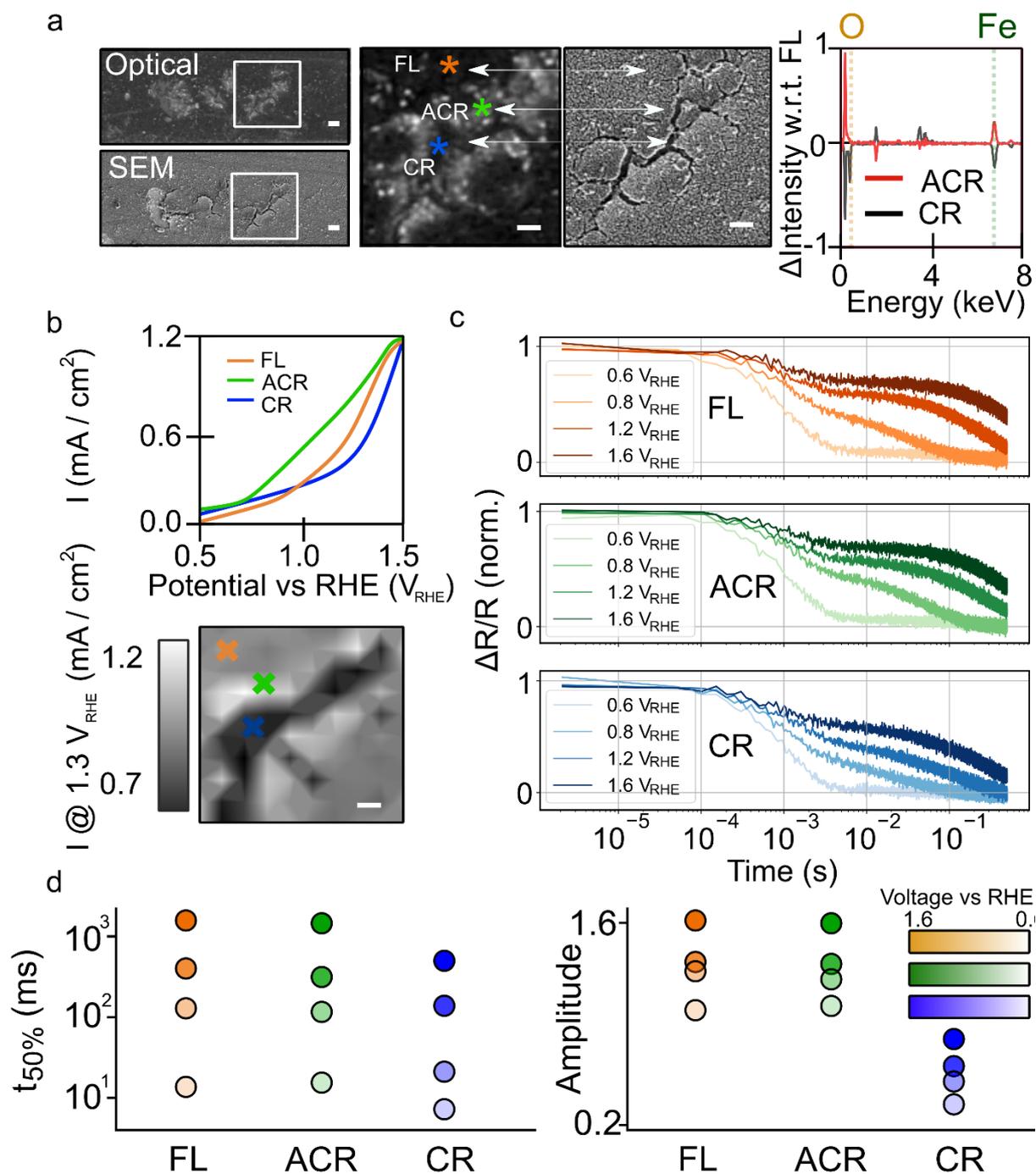

**Figure 3: OER reaction at and near cracks. a.** Optical and SEM image of prototypical region of hematite photoanode displaying cracks (blue asterisk), regions of α-$Fe_2O_3$ adjacent to crack (green/red asterisk) and untextured film (orange asterisk). Far right panel shows EDX spectra from adjacent to crack (ACR; red) and crack (CR; black) regions subtracted from the EDX spectrum in film (FL) regions. The CR regions show a depletion of both Fe and O whereas for ACR regions show an enhancement in only the Fe content; this suggest that these regions are oxygen deficient. **b.** Photocurrent density map at 1.3 $V_{RHE}$ of region shown in a. The orange, green and blue curves/crosses show CVs **c.** Kinetics of hole decay at 638 nm from FL, ACR and CR regions as a function of potential. Scale bar is 1 μm in all images. **d.** $t_{50\%}$ and amplitude of kinetics (see **Supplementary Note 4**) from FL, ACR and CR regions of sample. The error bars on values is not shown to ease visualisation, but there is a combined measurement and fitting error of ~10% on both $t_{50\%}$ and the amplitude.

We can use Raman imaging (**Figure 4c** and **4d**) to link the aforementioned kinetic and thermodynamic observations to local structure. The key Raman active modes in α-$Fe_2O_3$ are of $A^1_g$ and $E_g$ symmetry and are centred at 224 $cm^{-1}$ and 493 $cm^{-1}$ ($A^1_g$) and 244 $cm^{-1}$, 291 $cm^{-1}$, 408 $cm^{-1}$ and 610 $cm^{-1}$ ($E_g$). Structurally these correspond to the Fe-O bond stretch and the symmetric stretch of O-O (along the direction of Fe-O). Oxygen vacancies have been reported to lead to broadening of $E_g$ symmetry modes[50,51]. Additionally, there is a peak at 660 $cm^{-1}$ (and 1320 $cm^{-1}$ corresponding to double phonon or double magnon scattering). This mode (of $E_u$ symmetry) is assigned to FeOOH species and can be correlated with structural disorder/defects in the lattice or at grain boundaries[52,53]. The intensity of the Raman peaks report primarily on dynamic disorder (*i.e.* from phonon populations) whereas the peak positions and their widths probe static disorder (through changes in phonon lifetime as a result of crystal strain or vacancies)[54,55].

For the $E_g$ mode (and weakly for the $A^1_g$ modes) there is a similar intensity (same light blue colour) in the FL and CR regions as shown in the top row of **Figure 4d** (all measured at the open circuit potential (OCP)). However, for ACR regions the intensity of these modes is lower, suggesting lower dynamic disorder. A different trend is seen in the FWHM of the modes, where the $A^1_g$ modes are slightly broadened in the ACR region as compared to the FL/CR regions and the $E_u$ and $E_g$ modes whose widths specifically report on structural disorder narrow when making the same comparison; this supports the higher (oxygen) vacancy concentration in ACR. Under bias (and illumination) the Raman modes evolve in intensity and position. In ACR and CR, the $A^1_g$, $E_g$ and FeOOH modes all drop in intensity as the bias is raised, with $A^1_g$ and $E_g$ modes additionally shifting to higher wavenumbers (shift ~1.3 $cm^{-1}$; resolution = 0.1 $cm^{-1}$), as shown in the differential Raman spectra in **Figure 4c**. Interestingly, in FL regions the intensity of all modes increases under bias with a shift in the $A^1_g$ and $E_g$ modes to lower frequencies (~0.8 $cm^{-1}$). These changes with bias are challenging to rationalise, but suggest that there is a large reorganisation of vacancies and the formation of polarons, causing disorder in the structure, which change the length and electron density in Fe–O and O–O bonds as the bias is raised. Indeed, previous work[56] has suggested that water splitting in hematite occurs *via* the oxidation of a Fe(III)-O species to Fe(IV)=O, the formation of which would cause large lattice distortions (due to an accumulation of holes) and may be responsible for the bias dependent shifts.

More generally the changes in disorder with potential can also be observed in spatially resolved XANES measurements (2 μm spatial resolution; not the same film area as above). **Figure 4e, top left**, shows the principal component analysis (PCA) of XANES absorption maps of a 20 × 20 μm hematite surface, at 1.5 V vs RHE in dark conditions, at energy 7125 eV, *i.e.*, around the Fe-absorption edge. In these conditions the distribution of absorption values across the film does not show a strong spatial dependence (see **Supplementary Note 5** for further details). Following principle component decomposition (PCA), a cluster analysis can be subsequently done by grouping those pixels exhibiting statistically similar spectra between them. A cluster analysis from **Figure 4e, top left**, resulted in 5 clusters (blue, red, purple, green and orange data, respectively). Interestingly, upon turning on the illumination, *i.e.*, under OER conditions, the XANES absorption map (**Supplementary Note 5**) and its' PCA (**Figure 4e, top right**) of the same region, exhibit a spatial segregation of these clusters. Such cluster analysis after PCA is in broad agreement with the behaviours observed optically in **Figures 3** where we observe strong inhomogeneities in optoelectronic properties across the film under OER conditions. Fourier transforming the XANES data from the PCA analysis in R-space (**Figure 4e** bottom panel) shows that the relative intensity of the Fe–O bond (at 1.5 Å) between clusters, *i.e.*, different spatial positions, slightly differs (comparing green and red, for light off, and blue and orange when light is on). As discussed above, this is indicative of a different OV density distribution in agreement with the results shown in **Figure 3**. Additionally, the Fe–Fe bond length appears to exhibit a larger change in relative intensity in particular under illumination conditions, suggesting that it is specifically the degree of disorder that is spatially inhomogeneous, especially during OER conditions.

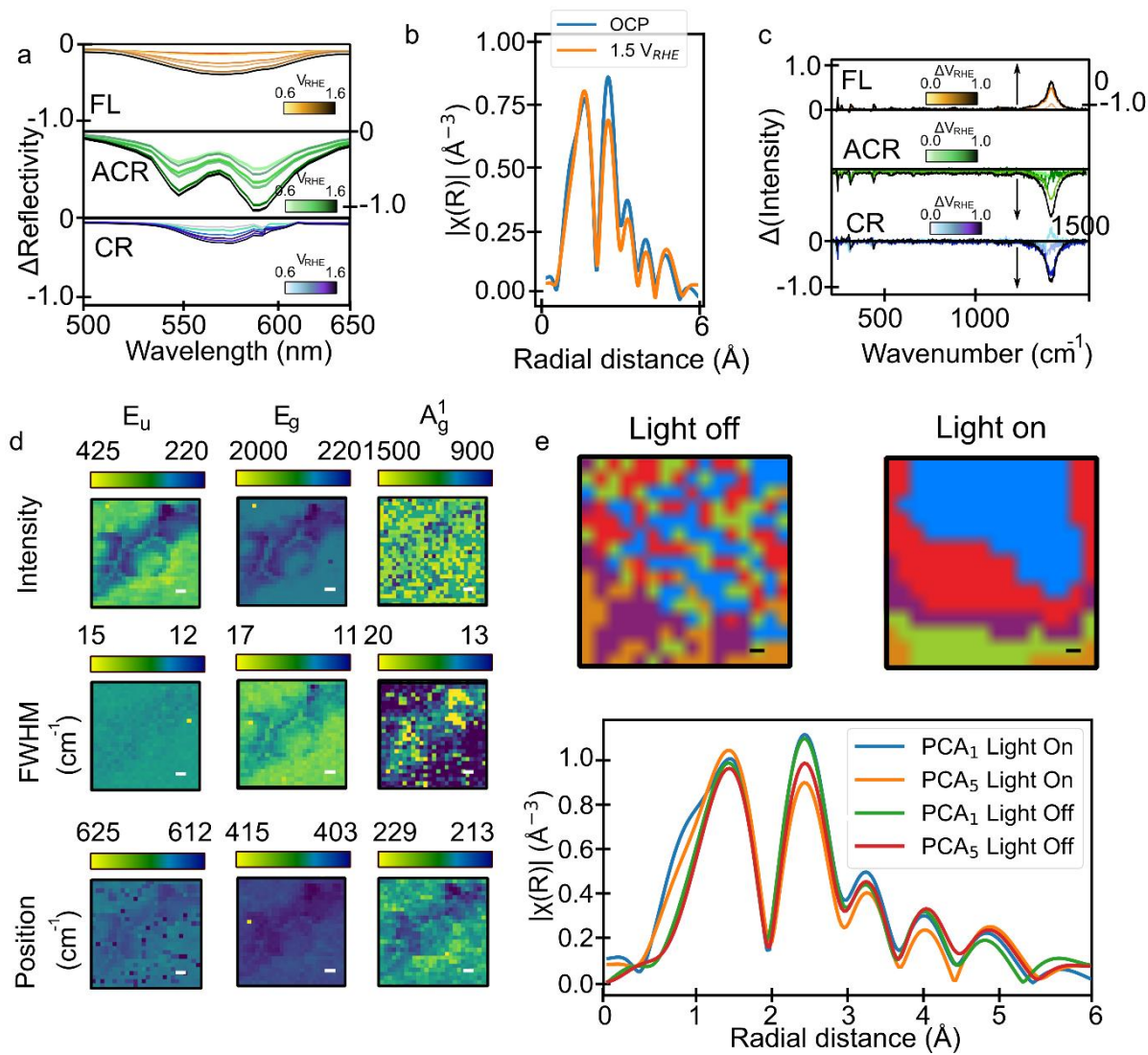

**Figure 4: Oxide vacancies and chemical structure at and near cracks. a.** Differential spectral reflectivity (with respect to reflection spectrum at 0.5 $V_{RHE}$) as a function of bias from the FL (oranges), ACR (greens) and CR (blues) regions marked in panel **a** of **Figure 3**. The differential reflectivity becomes more negative with bias for all regions showing a double negative peaked feature for ACR regions and a single negative peak for CR and FL regions. **b.** R-space plot of XANES spectra of the hematite (α-$Fe_2O_3$) photoanode in open circuit potential (*i.e,* no applied potential, blue trace), applying 1.5 V vs RHE (orange trace). **c.** Differential Raman spectra (with respect to 0.6 $V_{RHE}$) from region shown in **Figure 3a**. All scale bars are 1 um and panels show a 10 × 10 μm region. **d.** Image of intensity, FWHM and position of indicated Raman modes at OCP. **e.** (Top) Principal component analysis (PCA) of XANES absorption maps of a 20 × 20 μm hematite surface under dark (left) and under illumination (right) at 1.5 V vs RHE. The XANES absorption maps are found to contain 5 principle components denoted by the blue, red green, purple and orange colours in the maps. (Bottom) R-space plot of components 1 and 5 from of PCA analysis. Scale bar in images is 2 μm.

**The role of film thickness**

The thickness of a photoanode is also an important optimisation parameter. Too thin and absorption will be poor, but too thick and mobile charge carriers generated within the film will not be able to reach the electrolyte/catalyst surface interface due to the modest hole diffusion lengths in metal oxides. To assess this, we examine crack-free regions of the α-Fe$_2$O$_3$ film which present several different thicknesses as shown in **Figure 5a**. AFM topography scans reveal the three regions marked as T1, T2 and T3 with thicknesses of 310 ± 5 nm, 273 ± 8 nm and 145 ± 5 nm respectively (**Figure 5b**). SEM/EDX confirms that the percentage ratio Fe:O remains largely constant between the three layers as plotted in **Figure 5c**, *i.e.* we are only evaluating thickness effects. The content of Sn in T2 and T3 is ~3% lower than in T1, but given the 1 to 2 μm penetration depth of EDX such measurements may be dominated by the Sn signal from FTO. Nonetheless, we do not expect Sn diffusion into hematite layers to be significant. Interestingly, photocurrent density mapping, as shown in **Figure 5d**, shows that the photocurrent onset follows the order T3>T2>T1, *i.e.* potential for photocurrent generation and back electron transfer is lowest in T3. The photocurrent density in T3 at ~1.6 V$_{RHE}$ also is slightly larger (~1.77 mA cm$^{-2}$) than T1/T2 (~1.70 mA cm$^{-2}$) indicating that the fill-factor is largest in T3 (similar to rationale in above section).

Examining the hole population decay we find in T2 and T3, $t_{50\%}$ is systematically larger for a given bias than in T1 (**Figure 6a**) indicative of reduced BER in these regions (*e.g.* at 1.6 V$_{RHE}$ $t_{50\%}$ is ~1300 ms in T2/T3 and ~940 ms in T1). The magnitude of the ΔR/R signal also is ~1.4 times larger in T2/T3 as compared to T1 (see **Supplementary Note 6**) suggesting a larger hole population in-line with the higher photocurrent density at the saturating potential. Hole diffusion lengths are traditionally thought to be short[57] (10 to 100 nm) in metal oxide photocatalysts, hence despite recent reports of greater than 500 nm hole diffusion in Ti-doped hematite[58], we assume that diffusion properties of T1, T2 and T3 remain similar and are not responsible for our observations. One factor however that may be important is that thicker films are more defective (see discussion below), thus generating an excess of OVs that might act as trapping centres increasing recombination processes.

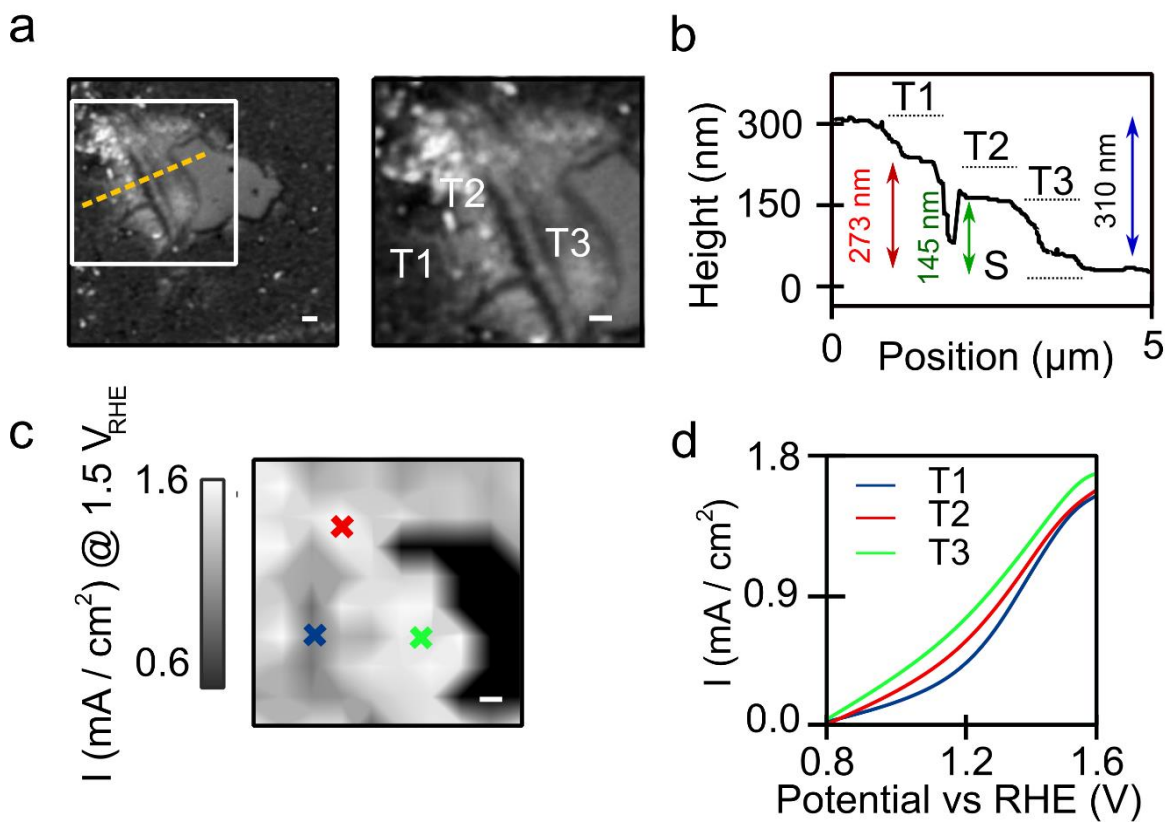

**Figure 5: Investigating the effect of photoanode thickness on activity. a.** Optical image of hematite photoanode displaying multiple thicknesses. The orange dashed line indicates where the AFM line scan is taken. Right panel shows 10 μm² region investigated here (white box in left panel). **b.** AFM topography profile with heights of three layers: T1, T2, T3 and substrate (S) marked. **c**. The Fe:O ratio appears constant between the three regions as obtained from EDX spectra. **d.** Photocurrent density map at 1.5 $V_{RHE}$ from corresponding area in panel **a**. Blue, red and green crosses/spectra taken from T1, T2 and T3 respectively. All scale bars are 1 um in the figure and panel sizes show a 10 × 10 μm region.

To assess the defect properties, specifically in terms of oxygen vacancies, we perform biased spectral reflectivity measurements. **Figure 6b** shows difference reflectivity spectra (*versus* 0.5 $V_{RHE}$) from T1, T2 and T3 on increasing the bias from 0.6 $V_{RHE}$ to 1.6 $V_{RHE}$. The spectra, as discussed above, show a broad negative feature centred between 560 nm and 580 nm whose magnitude increases with bias. In T1 there is a larger change in the magnitude of the reflectivity, on increasing the bias, than T2/T3. Based on the previous assignment of features (for ACR/FL/CR) this suggests that, in T1 there are a larger number of oxygen vacancies as compared to T2/T3 (the ratio of reflectivity T1:T2 at 570 nm is ~3.5). Although we cannot quantitatively extract the vacancy concentration, the low fill-factor in T1 would suggest that [OV] is sufficiently high that vacancies are in a regime where they act as detrimental trap sites as opposed to facilitating conduction (*i.e.* being beneficial). This also explains the shorter hole lifetimes in T1 due to defect-facilitated BER, along with the fact that electrons generated close to the surface need to travel further in the thicker parts of the film.

Finally, to assess other forms of structural disorder we repeat Raman imaging over the same region as shown in **Figure 5a**. Several features can be noted in the Raman images in **Figure 6c**: (i) the average intensity ratio $A^1_g$:$E_g$ is 1.34 in T1, whereas it is 1.02 and 0.93 in T2 and T3 respectively, (ii) the $A^1_g$ mode is shifted to lower frequencies in T2 and T3 and (iii) the FeOOH mode ($E_u$ symmetry) is also of higher intensity in T2 and T3 as compared to T1 suggesting there is some loss of symmetry and formation of tetragonal defects. Interestingly, these differences between T1 and T2 qualitatively resemble those when hematite is thinned to the few layer limit (hematene). In this case a high density of surface-active enhanced reaction sites and a locally modified charge around the adsorption sites[59,60] results in a greater photocatalytic activity with similar behaviour also reported for $BiVO_4$[61]. The results in **Figure 6c** suggest that similar structural modifications could be occurring in T2/T3 promoting water oxidation in these regions.

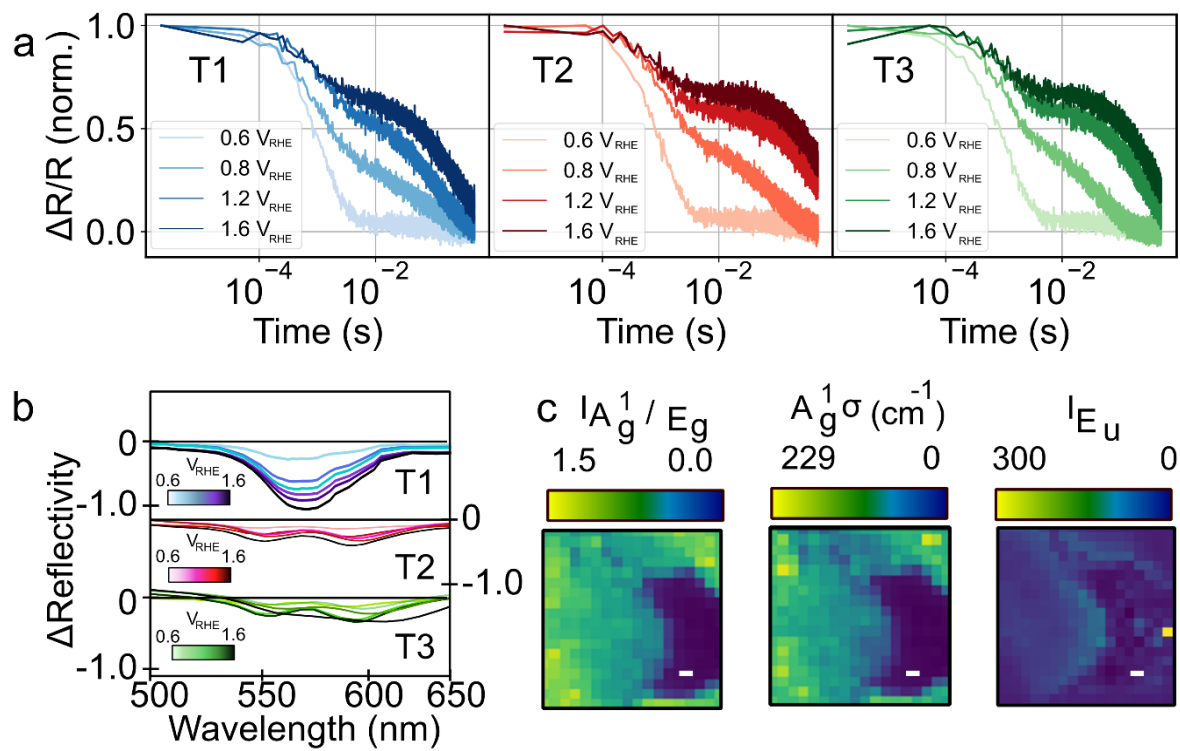

**Figure 6: Influence of thickness on hole dynamics, oxide vaccancies and local structure. a.** Kinetics of hole decay from T1, T2 and T3 regions (marked in **Figure 5a**) as a function of bias. **b.** Differential spectral reflectivity (with respect to reflection spectrum at 0.5 $V_{RHE}$) as a function of bias from T1, T2 and T3. **c.** Ratio of Raman mode intensities $A^1_g/E_g$ along with peak position (σ) for $A^1_g$ and intensity of the mode for $E_u$. When taking the ratio in the region of the substrate values are set to zero due to the high noise. All scale bars are 1 um in the figure and panel sizes show a 10 × 10 μm region.

Whilst we have focussed on mapping the influence of structural defects on (photo)catalytic activity, the above approach can also be used to examine the role of chemical impurities. In **Supplementary Note 7** we examine the influence of native carbon containing impurities on hematite photocatalysis. We find while these impurities increase the hole-lifetimes, the poor absorption and reduced oxygen vacancies they imbue makes them detrimental for photocatalysis.

**Conclusion**

We have demonstrated that microstructural and chemical defects strongly impact the 'local' J-V fill-factor (and hence efficiency) in an archetypal metal-oxide (photo)electrocatalyst, hematite. This is by modulating both the rates of back electron recombination and influencing the oxide vacancy concentration such that a balance between charge trapping and electronic conduction can be achieved. Regions of α-$Fe_2O_3$ adjacent to cracks and those that are thinned (~150 nm) relative to the bulk (~ 300 nm) show higher activity than the native film. More generally, our work highlights the need to move beyond macroscopic measurements and pay attention to microscopic heterogeneity when evaluating photoelectrochemical performance even in thin-film metal oxides. The all-optical approach shown here achieves this in-*situ*, label-free and non-destructively allowing a variety of measurements to be performed on the same location, as well as other complementary measurements such as SEM or XANES mapping.

With respect to defects in photocatalysts specifically, future studies may focus on examining emerging non-oxide systems such as organic photocatalysts, where the role of charge trapping in energy losses and reduction in reactivity is not fully understood[62]. Another case where microscopic resolution will be crucial is to better understand how the photoelectrode structure can be engineered such that polaron associated losses can be minimised[63] as well as monitoring the evolution and activity of microscale defects during long operation. In terms of the measurements, using an LED for excitation in pump-probe microscopy measurements will allow parameters to be extracted under even more relevant solar irradiance conditions[64]. Alongside this, other non-linear techniques, such as stimulated Raman scattering microscopy[65,66] will allow both faster acquisition of Raman images and improved optical resolution. The pump-probe approach we employ can be used to track the dynamics of a wide range of carriers. For example, the transient reflectance signal can also be related to the thermal diffusivity[67,68]. By mapping this across an electrode[69] or thermoelectric material[70], local heat transport properties could be estimated and correlated with structure.

**Methods**

Sample preparation

Undoped $\alpha$-$Fe_2O_3$ photoanodes were deposited by a hydrothermal method previously reported[37]. Briefly, irregular FeOOH was grown onto 25 mm × 25 mm FTO coated glass from a solution containing 150 mM $FeCl_3$ and 1 M $NaNO_3$ at 100 °C for 1h. After rinsing with DI water, the FeOOH was converted into $\alpha$-$Fe_2O_3$ by annealing at 500 °C for 1h.

Transient reflection microscopy

An EKSPLA NT340 tunable laser was used as the pump source. The pulse duration was 5 ns and repetition rate set to 0.1 Hz using a pulse picker. All experiments were performed at pump wavelengths between 380 nm and 510 nm; main text 400 nm. 3 O.D. 1.0 absorptive filters (Thorlabs) were used to directly attenuate the output of the laser. The collimated pump was pinholed (75 μm pinhole), telescoped (300:200 mm AB coating; Thorlabs) and directed using a 45 degree mirror to the objective (Olympus MPLAPON60X, 0.9 N.A) housed in a home-built microscope body. The probe was delivered by a 638 nm CW Coherent CUBE laser. The probe was pinholed (20 μm pinhole), passed through a 50:50 beam splitter (for collection) and combined with the pump line using a dichroic mirror (DLP505 Thorlabs). The probe was focussed into the back focal plane of the objective using the same 200 mm lens telescoping the pump. Samples were placed in a 3-electrode cell (MM Raman ECFC 3.5 $cm^2$, 4.5 mL; redox.me) with the housing modified to accommodate for the objective working distance. The sample cell was built onto a x-y-z translation (Optosigma) stage to allow for fine focussing and sample scanning. The reflected probe was collected by the beam splitter and then passed through a second 80:20 beam splitter, with 20% of the widefield probe imaged onto a camera (FLIR Grasshopper 3) and the remainder focussed with a 50 mm lens onto an avalanche photodiode (APD; Thorlabs); long pass filters were used to reject any scattered pump. The time resolution of the setup was found to be ~30 ns. The output of the APD was connected to an oscilloscope (Keysight Infium 9000; ns to μs) or National Instruments (NI USB-6211) DAQ card (μs to ms) for measurement of the kinetics depending on the time range. Data were stitched together between the two time ranges in post processing. 10000 laser shots were recorded per spatial location to achieve sufficient signal to noise. Potentials were applied with a Gamry Interface 5000E potentiostat. Home-built Python software was written for hardware interfacing and data acquisition. The probe intensity was set to ~100 μJ/$cm^2$ at the sample and the pump fluence varied to a maximum of ~500 µJ/cm$^2$ at the sample. Further details such as the wavelength and fluence dependence of signals is discussed in **Supplementary Note 8**.

Micro-spectroelectrochemisty

The same microscope body and cell used for TRM measurements was used for and micro-spectroelectrochemistry. No pump source was used, but instead a Xe white light (WL) lamp (200 W Newport) was passed through a monochromator (Bentham Instruments) and then coupled (Thorlabs RC02FC) into an optical fibre (500 µm diameter). The fibre output was focussed into the back focal plane of the same microscope objective as for TRM, to project the beam in the wide-field onto the sample. At a given bias, the monochromator wavelength was scanned in 3 nm steps and the reflection image measured on the camera (FLIR Grasshopper 3). Any beam splitters in the collection path were removed to collect the maximum reflected light. The bare FTO substrate was used to determine reference spectra subtracted from all data. Averaging over typically 25 spatial pixels allowed sufficient signal-to-noise to determine reflection spectra from the image stack.

Photo-electrochemical microscopy

Micro-photocurrent mapping was performed using the method outlined previously by Kozlowski *et al.*[71], Furtak *et al.*[72], Sambur *et al.*[73] and Butler[74]. Briefly, a 532 nm CW laser (Oxxius LCX-532S) was used to excite spots on the hematite anode through the back-side of the ITO. The laser focus diameter was ~1.3 µm with a maximal power density of 8 MW cm$^{-2}$. The 532 nm laser excitation light was chopped at 36 Hz and a lock-in amplifier (Stanford Instruments) was used to detect the current from the excited region of the photoanode on top of the background of the rest of the photoanode *via* a potentiostat (Gamry Interface 5000E). The steady state photocurrent signal was averaged for 30 s at each potential. The measured nanoampere photocurrents were normalized by the laser spot size to give a photocurrent density. The potential was stepped in 0.05 V increments from 0.5 $V_{RHE}$ to 1.6 $V_{RHE}$ with a smoothing spline interpolation between points to create the CV.

Raman microscopy

Raman measurements were performed with a Renishaw inVia Raman microscope in ambient condition. Excitation was provided with a 785 nm laser which is which is off-resonant from the hematite absorption ensuring no additional photogenerated species are created. The Raman emission was collected by a Leica 50× objective (N.A. = 0.85) and dispersed by a 1800 lines mm$^{-1}$ grating. The samples were scanned with an x-y piezo stage to collect 20000 spectra over 10 × 10 µm grid in 500 nm steps with 2 s dwell time at each spot. For potential and illumination dependent Raman measurements, the same configuration as for imaging was used with Raman spectra measured at each of the respective locations at a given potential. In this case spectra were acquired with a 3 s dwell time at each spot. A fibre white light source was used to illuminate the samples.

Atomic force microscopy

AFM measurements were done in tapping mode (Veeco Dimension 3100) at room temperature. The AFM cantilever was provided by MikroMasch. The tip radius was ∼10 nm.

SEM and EDX analysis

Secondary electron images were obtained using a MIRA3 TESCAN SEM system, at 5 kV. EDX spectra acquisition and analysis was performed using an Oxford Instruments AZtecEnergy X-Max$^N$ 80 EDX system, at 15 kV.

XANES

Both point and mapping XANES experiments were collected at the Fe K edge using a Si(111) double-crystal monochromator and the Xpress3 Si detector (Si D BL18I-EA-DET-02) in reflectance mode. The energy steps used were: 6962 to 7092 eV, 5 eV steps; 7092 to 7105 eV, 1 eV steps; 7105 7120 eV, 0.25

eV steps; 7120 7172 eV, 1 eV steps; 7172 7212 eV, 2 eV steps; 7212 7512 eV, 5 eV steps. A 0.05 mm Al filter was used to attenuate the fluorescence of the Fe K$\alpha$1 signal to avoid detector saturation. A sample stage (T1-BL18I-MO) was used for mapping purposes, performing coarse maps of 20 × 20 µm, with a 1 µm step for both x and y axes. An Fe foil was used for calibration purposes and $FeCl_2$, $FeCl_3$, $Fe_2O_3$ and $Fe_3O_4$ were used as model compounds for $Fe^{2+}$ and $Fe^{3+}$ species.

XANES data analysis was carried out using Athena Demeter 0.9.26 XAS data processing software and maps data were analysed by Mantis 2.3.02 software package[75].

**Acknowledgements**

R.P thanks Arjun Ashoka (Cambridge) for useful assistance and advice with building of the experimental setup and writing of the data acquisition code.

**Funding**
C.A.M. acknowledges the Generalitat Valenciana (APOSTD/2021/251 fellowship and MinCiencias Colombia through the Fondo Nacional de financiamiento para la ciencia, la tecnología y la innovación "Francisco José de Caldas", call 848 de 2019 for funding. A.J.M. acknowledges support from the EPSRC Cambridge NanoDTC, EP/L015978/1. R.P. acknowledges Clare College, Cambridge for funding *via* a Junior Research Fellowship. We acknowledge financial support from the EPSRC via grants EP/M006360/1 and EP/W017091/1 and the Winton Program for the Physics of Sustainability.

**Competing interests**
The authors declare that they have no competing interests.

**Data and materials availability**
The data that support the plots within this paper and other findings of this study are available at the University of Cambridge Repository (https://doi.org/XXXXX).

**Author Contributions**
C.A.M. measured and analysed X-ray data, interpreted the results and synthesised samples. M.S. interpreted the results. E.P. performed X-ray measurements and interpreted the results, N.B.G. performed Raman measurements. A.J.M. performed SEM and EDX measurements. M.A.G-G. interpreted the X-ray data. K.I. supervised X-ray measurements. S.G. supervised the work of C.A.M., A.R. supervised the work of A.J.M. and N.B.G.. J.D. interpreted the results and designed the project. R.P. designed the project, built and coded the transient optical microscope, performed measurements and analysed and interpreted the data. All authors contributed to the writing of the manuscript.

**Rights Retention Statement**